\documentclass[compsoc, conference, a4paper, 10pt, times]{IEEEtran}
\usepackage[T1]{fontenc}
\usepackage[inline]{enumitem}
\usepackage[latin9]{inputenc}
\usepackage[normalem]{ulem}
\usepackage{amsthm}
\usepackage{amsmath}
\usepackage{amssymb}
\usepackage{algpseudocode}
\usepackage{bm}
\usepackage{breakcites}
\usepackage{cite}
\usepackage{color}
\usepackage{colortbl}
\usepackage{float}
\usepackage{graphicx}
\usepackage{hhline}
\usepackage{listings}
\usepackage{mathrsfs}
\usepackage{multicol}
\usepackage{multirow}
\usepackage{url}
\usepackage{pgfplots}
\usepackage{stfloats}
\usepackage{subcaption}

\makeatletter

\floatstyle{ruled}
\newfloat{algorithm}{tbp}{loa}
\providecommand{\algorithmname}{Algorithm}
\floatname{algorithm}{\protect\algorithmname}
\usepackage{tikz}
\usetikzlibrary{calc}

\theoremstyle{plain}

\theoremstyle{plain}

\theoremstyle{plain}

\theoremstyle{plain}

\usepackage[acronym]{glossaries}
\newcommand{\newac}{\newacronym}
\newcommand{\ac}{\gls}
\newcommand{\Ac}{\Gls}
\newcommand{\acpl}{\glspl}

\newac{speb}{SPEB}{square position error bound}
\newac[plural=EFIMs,firstplural=Fisher information matrices (EFIMs)]{efim}{EFIM}{Fisher information matrix}
\newac{ne}{NE}{Nash equilibrium}
\newac{mse}{MSE}{mean squared error}
\newac{toa}{TOA}{time-of-arrival}
\newac{snr}{SNR}{signal-to-noise ratio}
\newac{lan}{LAN}{local area network}
\newac{psd}{PSD}{positive semidefinite}
\newac{pd}{PD}{positive definite}
\newac{wrt}{w.r.t.}{with respect to}
\newac{lhs}{L.H.S.}{left hand side}
\newac{wp1}{w.p.1}{with probability 1}
\newac{kkt}{KKT}{Karush-Kuhn-Tucker}
\newac{wlog}{w.l.o.g.}{without loss of generality}
\newac{mle}{MLE}{maximum likelihood estimation}
\newac{gps}{GPS}{global positioning system}
\newac{rssi}{RSSI}{received signal strength indicator}
\newac{mimo}{MIMO}{multiple-input multiple-output}
\newac{csi}{CSI}{channel state information}
\newac{fdd}{FDD}{frequency division duplexing}
\newac{ms}{MS}{mobile station}
\newac{bs}{BS}{base station}
\newac{d2d}{D2D}{device-to-device}
\newac{slnr}{SLNR}{signal-to-interference-leakage-and-noise-ratio}
\newac{ula}{ULA}{uniform linear antenna array}
\newac{pas}{PAS}{power angular spectrum}
\newac{mmse}{MMSE}{minimum mean square error}
\newac{zf}{ZF}{zero-forcing}
\newac{rzf}{RZF}{regularized zero-forcing}
\newac{as}{AS}{angular spread}
\newac{aod}{AOD}{angle of departure}
\newac{iid}{i.i.d.}{independent and identically distributed} 
\newac{sinr}{SINR}{signal-to-interference-and-noise ratio}
\newac{tdd}{TDD}{time-division duplex}
\newac{rvq}{RVQ}{random vector quantization}
\newac{rhs}{R.H.S.}{right hand side}
\newac{mrc}{MRC}{maximum ratio combining}
\newac{cdf}{CDF}{cumulative distribution function}
\newac{a.s.}{a.s.}{almost surely}
\newac{los}{LOS}{line-of-sight}
\newac{jsdm}{JSDM}{joint spatial division and multiplexing}
\newac{map}{MAP}{maximum a posteriori}
\newac{klt}{KLT}{Karhunen-Lo\`eve Transform}
\newac{lbe}{LBE}{link bargaining equilibrium}
\newac{se}{SE}{Stackelberg equilibrium}
\newac{uav}{UAV}{unmanned aerial vehicle}
\newac{nlos}{NLOS}{non-line-of-sight}
\newac{pdf}{PDF}{probability density function}
\newac{em}{EM}{expectation-maximization}
\newac{knn}{KNN}{$k$-nearest neighbor}
\newac{svd}{SVD}{singular value decomposition}
\newac{nmf}{NMF}{non-negative matrix factorization}
\newac{umf}{UMF}{unimodality-constrained matrix factorization}
\newac{rmse}{RMSE}{rooted mean squared error}
\newac{olos}{OLOS}{obstructed line-of-sight}
\newac{mmw}{mmW}{millimeter wave}
\newac{ber}{BER}{bit error rate}
\newac{rss}{RSS}{received signal strength}
\newac{lp}{LP}{linear program}
\newac{ufw}{U-FW}{unimodal Frank-Wolfe}
\newac{utf}{UTF}{unimodality-constrained tensor factorization}
\newac{fw}{FW}{Frank-Wolfe}
\newac{iot}{IoT}{Internet-of-Things}
\newac{mae}{MAE}{mean absolute error}
\newac{crb}{CRB}{Cram\'er-Rao bound}
\newac{aoa}{AoA}{angle of arrival}
\newac{wcl}{WCL}{weighted centroid localization}
\newac[plural=GNSS,firstplural=global navigation satellite systems (GNSS)]{gnss}{GNSS}{global navigation satellite system}
\newac{gsm}{GSM}{global system for mobile communications}
\newac{imu}{IMU}{inertial measurement unit}
\newac{rbf}{RBF}{radial basis function}
\newac{msf}{MSF}{multi-sensor fusion}
\newac{lidar}{LiDAR}{light detection and ranging}
\newac{glrt}{GLRT}{generalized likelihood ratio test}
\newac{sdr}{SDR}{software-defined radio}
\newac{ap}{AP}{access point}
\newac{lte}{LTE}{Long-Term Evolution}
\newac{raim}{RAIM}{receiver autonomous integrity monitoring}
\newac{pvt}{PVT}{position, velocity and time}
\newac{npl}{NPL}{Neyman-Pearson lemma}
\newac{ekf}{EKF}{extended Kalman filter}
\newac{tdoa}{TDOA}{time difference of arrival}
\newac{dop}{DOP}{dilution of precision}
\newac[plural=SOPs,firstplural=signals of opportunity (SOPs)]{sop}{SOP}{signal of opportunity}
\newac{ssid}{SSID}{service set identifier}
\newac{bssid}{BSSID}{basic service set identifier}
\newac{mac}{MAC}{medium access control}

\makeatother

\providecommand{\corollaryname}{Corollary}
\providecommand{\lemmaname}{Lemma}
\providecommand{\propositionname}{Proposition}
\providecommand{\theoremname}{Theorem}

\definecolor{mycolor1}{rgb}{0.494117647058824,0.184313725490196,0.556862745098039}
\definecolor{mycolor2}{rgb}{0.466666666666667,0.674509803921569,0.188235294117647}
\definecolor{mycolor3}{rgb}{0.301960784313725,0.745098039215686,0.933333333333333}
\definecolor{mycolor4}{rgb}{0.929411764705882,0.694117647058824,0.125490196078431}
\definecolor{mycolor5}{rgb}{0.635294117647059,0.078431372549020,0.184313725490196}
\definecolor{mycolor6}{rgb}{0.8500,0.3250,0.0980}
\pgfplotsset{
  every axis plot/.append style={line width=1.5pt},
}

\title{Position-based Rogue Access Point Detection}

\author{
    \IEEEauthorblockN{Wenjie Liu}
    \IEEEauthorblockA{
    \textit{Networked Systems Security Group}\\
    \textit{KTH Royal Institute of Technology}\\
    Stockholm, Sweden}
    wenjieli@kth.se
    \and
    \IEEEauthorblockN{Panos Papadimitratos}
    \IEEEauthorblockA{
    \textit{Networked Systems Security Group}\\
    \textit{KTH Royal Institute of Technology}\\
    Stockholm, Sweden}
    papadim@kth.se
}
\begin{document}

\maketitle              % typeset the header of the contribution
%

% biography section. The * indicates a section excluded from numbering.
% \section*{biography}

% Biographies are defined as follows:
% \biography{Author name}{author biography text}

% \biography{Wenjie Liu}{received the B.Eng. degree from the University of Electronic Science and Technology of China, Chengdu, China, in 2019, and the M.Phil. degree from The Chinese University of Hong Kong, Shenzhen, in 2021. He is currently pursuing the Ph.D. degree with the School of Electrical Engineering and Computer Science, KTH Royal Institute of Technology. He was a Visiting Student/an Exchange Student at Peking University and the Harbin Institute of Technology and interned at the Intel Asia-Pacific Research and Development Center. His research interests mainly include security and privacy.}

% \biography{Panos Papadimitratos}{is a professor with the School of Electrical Engineering and Computer Science (EECS) at KTH Royal Institute of Technology, Stockholm, Sweden, where he leads the Networked Systems Security (NSS) group. He earned his Ph.D. degree from Cornell University, Ithaca, New York, in 2005. His research agenda includes a gamut of security and privacy problems, with an emphasis on wireless networks. He is an IEEE Fellow, an ACM Distinguished Member, and a Fellow of the Young Academy of Europe.}

% The Abstract. The * indicates a section excluded from numbering.
\begin{abstract}
Rogue Wi-Fi \ac{ap} attacks can lead to data breaches and unauthorized access. Existing rogue \ac{ap} detection methods and tools often rely on \ac{csi} or \ac{rssi}, but they require specific hardware or achieve low detection accuracy. On the other hand, \ac{ap} positions are typically fixed, and Wi-Fi can support indoor positioning of user devices. Based on this position information, the mobile platform can check if one (or more) \ac{ap} in range is rogue. The inclusion of a rogue \ac{ap} would in principle result in a wrong estimated position. Thus, the idea to use different subsets of \acpl{ap}: the positions computed based on subsets that include a rogue \ac{ap} will be significantly different from those that do not. Our scheme contains two components: subset generation and position validation. First, we generate subsets of \acpl{rssi} from \acpl{ap}, which are then utilized for positioning, similar to \ac{raim}. Second, the position estimates, along with uncertainties, are combined into a Gaussian mixture, to check for inconsistencies by evaluating the overlap of the Gaussian components. Our comparative analysis, conducted on a real-world dataset with three types of attacks and synthetic \acpl{rssi} integrated, demonstrates a substantial improvement in rogue \ac{ap} detection accuracy. 
\end{abstract}

% \glsresetall

\begin{figure*}
    \centering
    \includegraphics[width=0.8\textwidth]{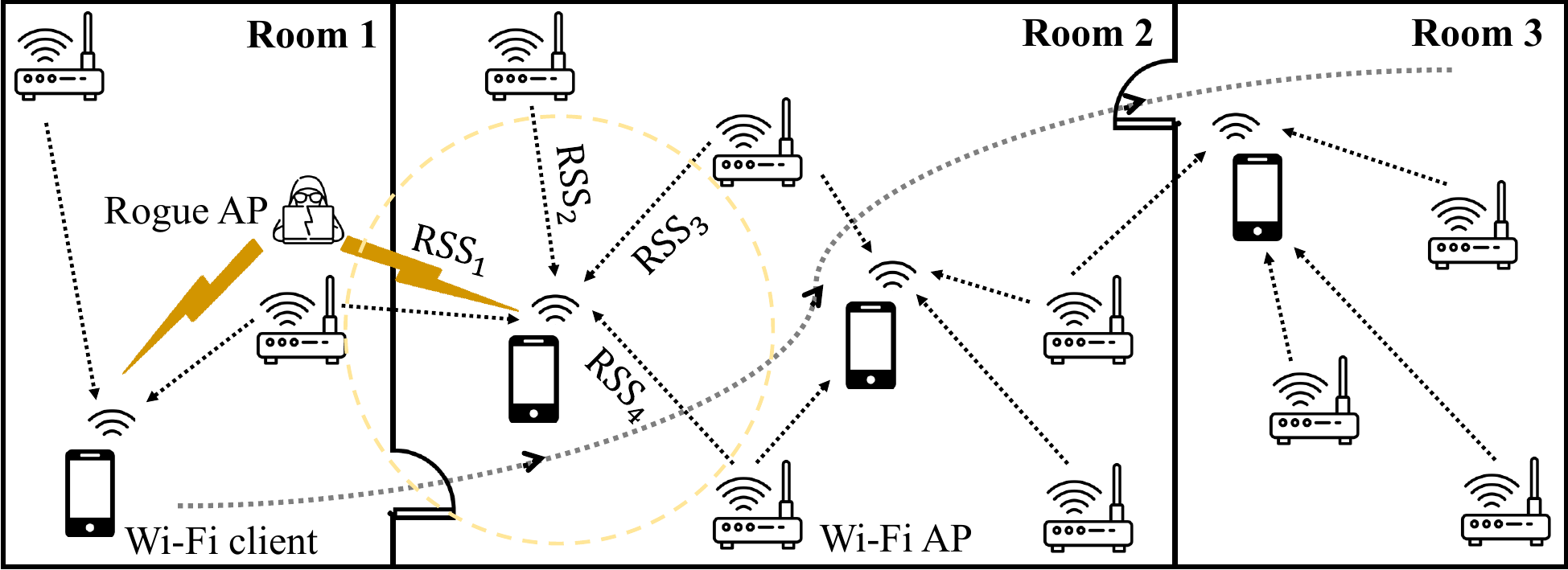}
    \caption{System and adversary model. For example, the mobile Wi-Fi client in the dashed yellow circle can locate itself using the subsets of four \acpl{rssi} from \acpl{ap} in range; the subset containing the rogue \ac{ap} would be inconsistent.}
    \label{fig:sysadv}
\end{figure*}

\section{Introduction}
Wi-Fi \acpl{ap} enable devices to connect to local area networks and the Internet wirelessly. However, rogue Wi-Fi \acpl{ap}, unauthorized \acpl{ap} installed in the area of a network without approval, pose significant cybersecurity risks. Rogue \acpl{ap} facilitate man-in-the-middle attacks (intercepting and altering communication of devices, attracted to connect to them instead of legitimate \acpl{ap}); or they are used in phishing campaigns (unsuspecting users connect to the rogue \ac{ap}, unknowingly reveal login credentials and other sensitive information) \cite{AloEll:J16}. Hence, considering the ease of deploying rogue \acpl{ap} and given most of the public \acpl{ap} are open networks, it is vital to have effective and robust rogue \ac{ap} detection. 

Existing solutions for rogue \ac{ap} detection, \cite{Aru:J24,Ibm:J24,Gan:J24}, simply use a whitelist of Wi-Fi \acpl{ap}. Other detection methods leverage techniques based on \ac{csi}, \ac{rssi}, or their combination. They identify unauthorized \acpl{ap} by analyzing signal characteristics or monitoring changes in the wireless environment. \cite{AhmAmiKanAbd:C14} uses \ac{rssi} and piggybacks information in IEEE 802.11 beacon frames to localize the Wi-Fi \ac{ap} and support location verification, but it modifies Wi-Fi. \cite{WuGuDonShi:J18} establishes feature vectors using \acpl{rssi} and then performs a clustering analysis on them, albeit with low detection accuracy. Location, hardware, and environment-related fingerprints based on \ac{csi} are explored in \cite{BagRoeMarSch:C15,LinGaoLiDon:C20,YanYanYanSon:J22}. However, these solutions require specialized hardware, complex configuration, or extensive computational resources, which limit practicality \cite{AloEll:J16}. Furthermore, dynamic network environments, as \ac{csi} is sensitive to environmental changes \cite{MaZhoWan:J19}, lead to false positives or missed detections. 

Detecting a rogue \ac{ap} that mimics a legitimate \ac{ap} (e.g., Evil-twin \cite{AloEll:J16,VanPie:C14,VanBhaDerOuz:C18}) is challenging and an active detection scheme, accessible to network operators and clients/users, is needed. Attackers can replicate hardware information from legitimate signals, allowing them to evade detection by hardware or signal fingerprinting methods. On the other hand, Wi-Fi \ac{rssi}-based indoor positioning is already very mature for daily use, with meter-level accuracy \cite{ZheLiLiaXue:J23}. This motivates the following question: can a position-based method, leveraging the deployment of multiple \acpl{ap}, typically so in many environments nowadays, detect rogue \acpl{ap}? This paper answers positively, inspired by the \ac{raim} method, developed for \ac{gnss} spoofing detection \cite{LiuPap:C24}, detecting rogue Wi-Fi \acpl{ap} based on positioning results and their cross-validation. Based on that, rogue \ac{ap} can be detected and excluded, relying solely on \acpl{rssi}.

The scheme we propose can work with contemporary mobile devices and mainstream Wi-Fi cards. The approach is intuitive: with access to a positioning algorithm and \ac{ap}-specific data, the mobile localizes itself. If it does so while taking into account a rogue \ac{ap}, the computed device position will be different from the one(s) computed with benign \acpl{ap} only. The mobile device uses \ac{rssi} subsets of all available \acpl{ap} and detects the attack through the positions that deviate when one (or more) rogue \acpl{ap} are part of a subset.

Our rogue \ac{ap} detection is compatible with many Wi-Fi positioning algorithms as long as they provide position and uncertainty, e.g., \ac{rssi} fingerprint-based positioning (based on a database \cite{ZheLiLiaXue:J23}, not hardware fingerprints), or distance-based positioning \cite{LiuChe:C21,Mozilla2023} (based on the positions of the deployed \acpl{ap}). Given our method is built on state-of-the-art positioning algorithms \cite{ZheLiLiaXue:J23,Mozilla2023}, it inherits advantages, such as insensitivity to environmental change. By analyzing the positioning results obtained from different \ac{ap} \ac{rssi} subsets, the method identifies positions consistent across multiple subsets. Inconsistency implies that a rogue \acpl{ap} is part of one or more subsets. For example, in Fig.~\ref{fig:sysadv}, four \acpl{ap} are within range of the Wi-Fi client entering (on the left of) Room 2: for a subset of two legitimate and one rogue \ac{ap}, the client position estimation differs from that involving any three benign \acpl{ap} in Room 2. The inconsistency reveals an attack and an examination of a sufficient number of subsets (positions) can reveal which is the rogue \ac{ap}, and then have it excluded (or even localized).

Our key contribution is a scheme that can work either as a tool used by wireless network operators for actively monitoring \acpl{ap} or by mobile users that can independently detect rogue \acpl{ap}. We just need \acpl{rssi} to make the scheme compatible with other rogue \ac{ap} detection schemes. We adopt and adapt the \ac{raim} used in \ac{gnss} to achieve rogue Wi-Fi \ac{ap} detection, and we term this Gaussian mixture \ac{raim}. We exploit position fusion from Wi-Fi positioning algorithms and rogue \ac{ap} exclusion strategies. Our numerical experiments and a comparative analysis on a partially real-world Wi-Fi \ac{rssi} dataset show significant improvement over baseline rogue \ac{ap} detection schemes. 

\section{Related Work}
\subsection{Rogue Wi-Fi AP Detection}
Rogue Wi-Fi \ac{ap} detection has received significant attention. Some industrial solutions \cite{Aru:J24,Ibm:J24,Gan:J24} ignore unknown \acpl{ap} or use a whitelist of \ac{ap} \acpl{mac} and \acpl{ssid} to find unauthorized ones; however, attackers can relatively easily forge this information. For example, many commercial Wi-Fi routers can set any \ac{ssid}, while open-source router systems, such as OpenWrt, can even modify their \ac{mac} address. The method in \cite{AhmAmiKanAbd:C14} uses \ac{rssi} measurements to identify rogue \acpl{ap}, but its clustering and two-step algorithm is affected by signal variations due to environmental factors (interference and multipath effects). Another approach employs wireless fingerprinting techniques \cite{LinGaoLiDon:C20}, independent of client devices; however fingerprinting scalability and robustness in dynamic network environments (rainfall, high pedestrian traffic, etc.) is challenging. PRAPD \cite{WuGuDonShi:J18} introduces a method based on \ac{rssi} for practical rogue \ac{ap} detection. Real-time identification of rogue Wi-Fi connections in operational networks was explored in \cite{YanYanYanSon:J22}. Additionally, \ac{csi} based on environment-related semantics for \ac{iot} \cite{BagRoeMarSch:C15}, while offering potential accuracy advantages, requires specialized hardware and large-scale scanning of frequency bands. 

\subsection{RAIM for GNSS Spoofing}
Receiver autonomous integrity monitoring (\Ac{raim}) leverages redundant data and performs consistency checks based on subsets of visible satellites \cite{Bro:J92}. Over the past few decades, there are mainly two types of \acpl{raim}: residual-based and solution separation \cite{JoeChaPer:J14}. Residual-based \ac{raim} \cite{KhaRosLanCha:C14,RoyFar:C17} uses statistical hypothesis testing on residual errors, identifying potential faulty measurements. The residual can be from least-squares or filters: \ac{ekf}-combined \ac{raim} utilizes rolling window filters to identify and eliminate outliers using \ac{gps} and inertial sensors \cite{KhaRosLanCha:C14,RoyFar:C17}. Solution separation \ac{raim} \cite{ZhaPap:J21,LiuPap:C24} recursively assumes faulty satellites, generates subsets of the remaining satellites to derive solutions, and then identifies which subsets contain the faults. For example, \cite{ZhaPap:J21} integrates RANSAC clustering to classify position solutions. Additionally, advanced \ac{raim} \cite{BlaWalEngLee:J15} extends fault exclusion to include multiple constellations like \ac{gps} and Galileo, providing enhanced security beyond \ac{gps}. Recent \ac{sop} techniques \cite{MaaKas:J21,LiuPap:C24} leverage wireless network signals for \ac{raim}, integrating kinematics models, cellular pseudoranges, or Wi-Fi measurements to improve \ac{raim} performance.

\section{System Model and Adversary}
% This section introduces the system model and the adversary addressed in this paper and establishes a notation system for this model.
\subsection{System Model}
We consider a mobile Wi-Fi device shown in Fig.~\ref{fig:sysadv}, with unknown position $\mathbf{p}_{\text{c}}(t) \in \mathbb{R}^3$ (coordinates and height)---the device can be held by a security patrol person from the network provider or be a user device (e.g., a smartphone). The client actively explores Wi-Fi \acpl{ap} and gathers beacon information from them. Then, $\mathrm{RSS}_j(t),j \in \mathcal{J}(t),t=1,2,...,T$, is \ac{rssi} information at time $t$, where $\mathcal{J}(t)$ is the set of \acpl{ap} identifiers at time $t$ providing \acpl{rssi}, and $T$ is the last time index. 

Although the client is indoors, walking and without \ac{gnss} reception, we assume pre-surveyed data, depending on the requirements of the Wi-Fi positioning algorithm. The data can be: (i) a fingerprint database of \acpl{rssi}, $\mathcal{T}=\left\{ \{\mathrm{RSS}_j(t) \mid j \in \mathcal{J}(t)\},\mathbf{p}_{\text{c}}(t) \right\}_{t=T_1}^{T_2}$, where $t=T_1,...,T_2$ are the time indexes for the database, or (ii) a list of \ac{ap} positions, $\mathbf{p}_{\text{AP}}^{(j)}(t) \in \mathbb{R}^3$ for $\mathrm{RSS}_j(t)$. 

\subsection{Adversary}
The rogue \acpl{ap}, often Evil-twin \acpl{ap} of legitimate ones \cite{VanPie:C14,VanBhaDerOuz:C18}, emulate the hardware model, \ac{ssid}, \ac{bssid}, and \ac{mac} address of legitimate \acpl{ap}, deceiving users (Wi-Fi clients) to connect. The attacker may employ a ``coexistence strategy'': if the \ac{rssi} of the legitimate \ac{ap} is relatively weak, the rogue \ac{ap} can boost its strength to compel clients to connect as well as relay packets to/from the legitimate \ac{ap}.\footnote{If the rogue \ac{ap} replaces or is very close to the legitimate ones, the security patrol person can easily find it without relying on our method \cite{AloEll:J16}.}

However, the attacker can not remove legitimate signals or completely mimic them. Hence, we assume that while rogue \acpl{ap} join the environment of in the same nearby legitimate ones to lure unsuspecting users, they inevitably change signal properties (\ac{rssi}) and differ in positions, thus leading to deviations of indoor Wi-Fi positioning results. 

\begin{figure*}
    \centering
    \includegraphics[width=0.8\textwidth]{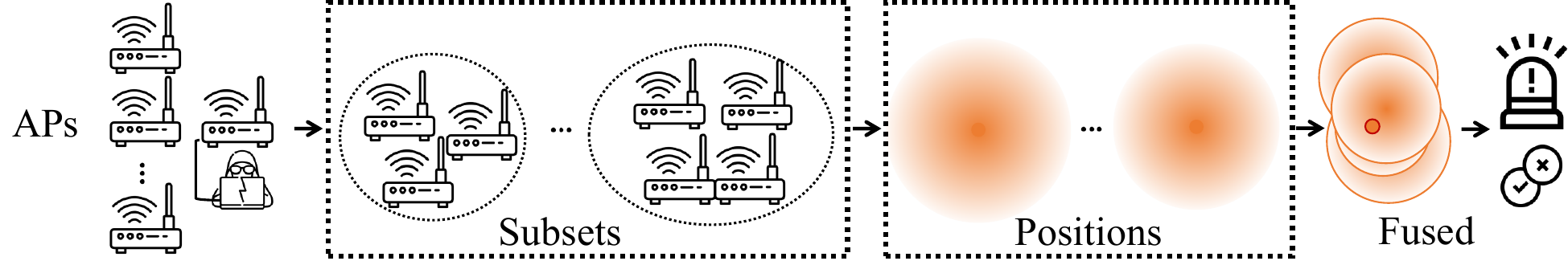}
    \caption{System overview of position-based rogue \ac{ap} detection. }
    \label{fig:scheme}
\end{figure*}

\section{Proposed Scheme}
\label{prosch}
% This section begins by presenting an overview of the proposed scheme, followed by introducing its two main detection algorithm components.
% The overall procedure of the proposed scheme is depicted in Fig.~\ref{fig:scheme}. 
\subsection{Subset Generation}
Subsets are generated as illustrated in Fig.~\ref{fig:scheme}, without presuming a known number of rogue \acpl{ap}. We systematically generate subsets considering all possible combinations of \acpl{ap}, varying in size---from the minimum to the maximum allowable for the given positioning algorithm. These subsets of \ac{rssi} indexes, $j$, are denoted as $\mathcal{S}_l(t)$, where $l=1,2,...,L(t)$ and $L(t)$ is the number of all subsets. An analysis of \ac{raim} resilience in \cite{LiuPap:C24} (single infrastructure case) shows that position estimates from benign subsets are around the actual position, and the number of them is most likely larger than the number of the attack subsets. 

\subsubsection{Sampling Strategy for Subsets}
\label{subsec:samstr}
The generation of subsets is a critical step for rogue \ac{ap} detection. Given the number of subsets can be large, we employ a straightforward sampling strategy: a random selection of subsets, with each subset chosen based on a predetermined probability distribution. The rationale is to select a diverse range of subsets without introducing significant bias or skewness. Despite its simplicity, this approach is theoretically robust with negligible impact on the cross-validation process, ensuring adaptability to various network configurations and attack scenarios. Most importantly, it enhances computation efficiency of our method. 

\subsubsection{Fingerprint-based Positioning}
It is necessary to construct a database of fingerprint vectors of \acpl{rssi} associated with the client positions in advance. Then, the positioning algorithm compares the current \ac{rssi} vector with the database and returns the estimated latitude and longitude coordinates and height of the receiver.

Given a subset that needs to be used for positioning, $\{\mathrm{RSS}_j(t) \mid j \in \mathcal{S}_l(t)\}$, and the fingerprint database $\mathcal{T}=\left\{ \{\mathrm{RSS}_j(t) \mid j \in \mathcal{J}(t)\},\mathbf{p}_{\text{c}}(t) \right\}_{t=T_1}^{T_2}$, which is pre-surveyed and without adversarial data, we look for the most $K$ similar fingerprints in the database $\mathcal{T}$. The function for scoring similarity to the fingerprint at $t'$ is defined as:
\begin{multline}
    f\left(\{\mathrm{RSS}_{j}(t)\mid j\in\mathcal{S}_{l}(t)\},\{\mathrm{RSS}_j(t') \mid j \in \mathcal{J}(t')\}\right)\\
    =\sum_{j \in \mathcal{J}(t)}\frac{\mathbb{I}\left\{ j \in \mathcal{S}_l(t) \right\} }{\max(\left|\mathrm{RSS}_{j}(t)-\mathrm{RSS}_{j}(t')\right|,d_{\mathrm{min}})}
\end{multline}
where $\mathbb{I}\{\mathrm{A}\}$ is 1 when the condition A is satisfied, and $d_{\mathrm{min}} > 0$ is the minimal difference of two \acpl{rssi}. Suppose the client position associated with the most $K$ similar fingerprints are $\mathbf{p}_{\text{c}}^{(k)},k=1,2,...,K$, with similarity scores $f^{(k)},k=1,2,...,K$. Then, the positioning result of $\hat{\mathbf{p}}_l(t)$ is the weighted average:
\begin{equation}
    \hat{\mathbf{p}}_l(t)=\frac{ \sum_{k=1}^K f^{(k)} \mathbf{p}_{\text{c}}^{(k)} }{ \sum_{k=1}^K f^{(k)} }
\end{equation}
with the reciprocal of the average of $\left\{f^{(k)}\right\}_{k=1}^K$ as the uncertainty of positioning, $\hat{\boldsymbol{\sigma}}_l(t)$.

\subsubsection{Distance-based Positioning}
Geolocation localization \cite{Mozilla2023} proposes a weighted nonlinear least squares problem to minimize the weighted sum of squared distances between \acpl{ap} and the estimated location. These weights are determined based on the inverse square of the signal strengths: 
\begin{equation}
    \underset{\hat{\mathbf{p}}_l(t)}{\mathop{\min}}\quad \sum_{j} \left(\frac{||\mathbf{p}_{\text{AP}}^{(j)}(t)-\hat{\mathbf{p}}_l(t)||_2}{\mathrm{RSS}_j(t)}\right)^2.
\end{equation}

Then, the positioning result and the least squares residual (position uncertainty) from $l$th subset at time $t$ are $\hat{\mathbf{p}}_l(t)$ and $\hat{\boldsymbol{\sigma}}_l(t)$, where $l(t)=1,2,...,L(t)$.

Apart from the fingerprint and distance-based methods, our subset generation is compatible with other Wi-Fi positioning algorithms providing position and uncertainty.

\subsection{Position Validation}
We see the positioning results as random variables following $\mathcal{N}(\hat{\mathbf{p}}_l(t),\hat{\boldsymbol{\sigma}}_l(t))$. Then, the Gaussian mixture at time $t$ is fused into a temporary position, $\tilde{\mathbf{p}}(t)$. Based on $\tilde{\mathbf{p}}(t)$, rogue \ac{ap} detection and exclusion are carried out. Furthermore, an ultimate position $\hat{\mathbf{p}}(t)$ can be estimated based on the \acpl{ap} after exclusion.

The temporary position fused from the Gaussian mixture is:
\begin{equation}
    % \Tilde{\mathbf{p}}(t)=\frac{1}{L(t)} \hat{\mathbf{p}}_l(t) \circ \hat{\boldsymbol{\sigma}}_l(t)
    % \tilde{\mathbf{p}}(t)=\frac{1}{L(t)}\hat{\mathbf{p}}_{l}(t)\circ\hat{\boldsymbol{\sigma}}_{l}(t)
    % \tilde{\mathbf{p}}(t)=\frac{\sum_{l=1}^{L(t)}\hat{\mathbf{p}}_{l}(t)\circ\hat{\boldsymbol{\sigma}}_{l}(t)}{\sum_{l=1}^{L(t)}\hat{\boldsymbol{\sigma}}_{l}(t)}
    \tilde{\mathbf{p}}(t)=\left(\sum_{l=1}^{L(t)}\hat{\mathbf{p}}_{l}(t)\oslash\hat{\boldsymbol{\sigma}}_{l}(t)\right) \oslash \left({\sum_{l=1}^{L(t)}\mathbf{1}\oslash\hat{\boldsymbol{\sigma}}_{l}(t)}\right)
\end{equation}
where $\oslash$ is the Hadamard division. 
% i.e., $[\hat{\mathbf{p}}_l(t) \circ \hat{\boldsymbol{\sigma}}_l(t)]_{i}=[\hat{\mathbf{p}}_l(t)]_i [\hat{\boldsymbol{\sigma}}_l(t)]_i$. 
Then, the deviation between the $l$th positioning result and the fused position at time $t$ is:
\begin{equation}
    d_l(t)=\lVert \hat{\mathbf{p}}_l(t) - \tilde{\mathbf{p}}(t) \rVert.
\end{equation}
If the deviation is larger than a threshold $\Lambda$, the subset for this positioning result is identified as a set containing rogue \ac{ap}. The threshold is:
\begin{equation}
    \Lambda = \mathbb{E}[d_l(t)] + n_\Lambda \cdot \sqrt{\mathbb{V}[d_l(t)]}
\end{equation}
where $n_\Lambda$ is a factor controlling coverage, usually taking a value of 3, according to the three-sigma rule of thumb. In most cases, larger $n_\Lambda$ comes with a higher false positive rate. Hence, by adjusting $n_\Lambda$, we can get the desired false positive and true positive rate. 

\subsubsection{Rogue AP Exclusion}
After validating all subsets $\mathcal{S}_l(t),l=1,2,...,L(t)$, the detected non-benign subsets can find the intersection or vote for the rogue \acpl{ap}. 

Denote the set of the index $l$ of non-benign subsets as $\mathcal{L}(t)$. Then, the identifiers of rogue \acpl{ap} at $t$ are:
\begin{equation}
    \bigcap_{l \in \mathcal{L}(t)} \mathcal{S}_l(t).
\end{equation}
However, considering an environment with positioning noise, we use the following voting scheme for classifying the $j$th \ac{ap} in $\mathcal{J}(t)$ as rogue or not:
\begin{equation}
    \begin{cases}
    \text{A} = \sum_{l \in \mathcal{L}(t)} \mathbb{I}\{j \in \mathcal{S}_l(t)\} \\
    \text{B} = \sum_{l \notin \mathcal{L}(t)} \mathbb{I}\{j \in \mathcal{S}_l(t)\}
    \end{cases}.
\end{equation}
If $\text{A}>\text{B}$, the $j$th \ac{ap} is rogue at time $t$. Then, $\hat{\mathbf{p}}(t)$ is calculated based on the remaining \acpl{ap}. 

\section{Experiments}
\label{experi}
\subsection{Dataset}
The dataset from \cite{ZheLiLiaXue:J23} was collected on the fourth floor of a shopping mall, covering an area of 170 by 90 m. 8 \acpl{ap} are visible at each position, while the landmarks and \ac{ap} positions were used to generate trajectories. The device WHUWearTrack was employed for trajectory acquisition as ground truth, incorporating a low-cost \ac{imu}, a Bluetooth module, and a multiprotocol system-on-chip. The reference trajectories have decimeter-level positioning accuracy. The Wi-Fi positioning algorithm from \cite{ZheLiLiaXue:J23} exhibits meter-level accuracy. 

\begin{figure}
    \centering
    \includegraphics[width=1.03\columnwidth]{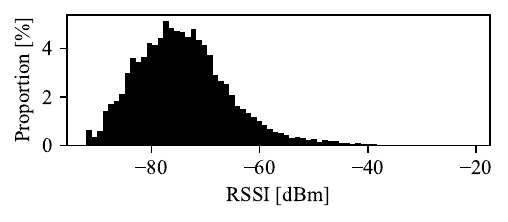}
    \caption{The distribution of the \ac{rssi} dataset.}
    \label{fig:rssdis}
\end{figure}

\subsubsection{Attack Simulation}
\label{subsec:attsim}
We generate rogue \ac{ap} signals by modifying the \ac{rssi} data: (i) An \ac{ap} is randomly chosen to have $\mathcal{N}(\mu_\mathrm{adv},\sigma_\mathrm{adv}^2)$ dB additive signal strength in a random duration (around one-third of the entire trace), where $\mu_\mathrm{adv}=10,\sigma_\mathrm{adv} \in \{2,4\}$, (ii) \acpl{rssi} of an \ac{ap} are replaced by values following $\mathcal {U}_{[-70,-55]}$ in dBm, or (iii) A rogue \ac{ap} located at $(30.52868,114.35086)$ and 12 meters in height, with its \acpl{rssi} simulated based on an indoor path loss model \cite{LiuPap:C23}. 16 traces containing rogue \ac{ap} signals for each attack are generated, so 48 in total. The distribution of \acpl{rssi} containing attacks is depicted in Fig.~\ref{fig:rssdis}. 
% \begin{table}
% \centering
% \caption{The \ac{rssi} distribution of the datasets.}
% \begin{tabular}{llllll}
% \hline
% \multirow{2}{*}{\ac{rssi} [dBm]} & Unusable & Bad & Okay & Good & Very Good \\ \cline{2-6} 
%                   & <-90 & -90~-80 & -80~-70 & -70~-67 & >-67 \\ \hline \hline
%                  Proportion [\%] &  &  &  &  &  \\ \hline
% \end{tabular}
% \label{tab:rssdis}
% \end{table}

\subsection{Baseline Methods}
The first baseline method is a clustering-based method inspired by \cite{WuGuDonShi:J18}, which uses a 2-class clustering algorithm to process \ac{rssi} of individual \ac{ap}. The $k$-medoid clustering algorithm has a distance measurement method that can deal with missing values and noise. If the distance between clusters is larger than a threshold, the corresponding \ac{ap} is classified as rogue. 

Another baseline is an anomaly detection scheme, ECOD~\cite{LiZhaHuBot:J22}, calculating the differences of \acpl{rssi} over $t$ per \ac{ap}. It estimates the underlying distribution of the input data through empirical cumulative distributions for each \ac{ap}, without relying on predetermined parameters. These estimates serve as the basis for determining tail probabilities associated with each data vector across all \acpl{ap}. Last, ECOD derives an outlier score by aggregating the estimated tail probabilities across \acpl{ap} at $t$. 

We use true positive rate, $P_\text{TP}$, as the metric for comparison, defined as the number of true positive detections (or exclusions) of rogue \acpl{ap} over the total number of positive (rogue \ac{ap} launching attacks) timestamps. To compare the methods fairly, we fix the false positive rate, $P_{\text{FP}}$, to $0.01,0.02,...,0.1$. A false positive is a detection result that wrongly believes a rogue \ac{ap} attack takes place while it does not. 

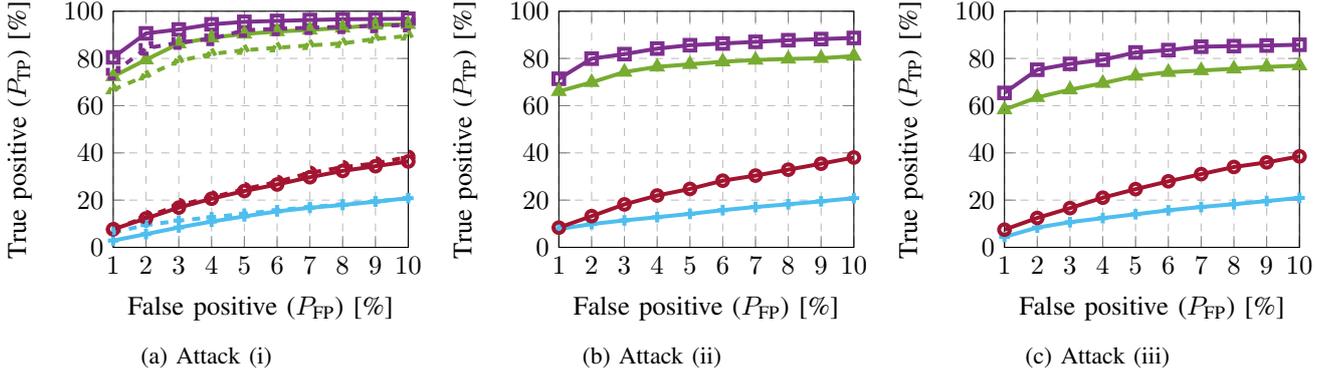
\begin{figure*}
    \centering
    \begin{subfigure}[b]{0.66\columnwidth}
    \centering
    \begin{tikzpicture}%[scale=.75,font=\large]
    \begin{axis}[
        width=\columnwidth,
        xlabel={False positive rate ($P_{\text{FP}}$) [\%]},
        ylabel={True positive rate ($P_\text{TP}$) [\%]},
        xmin=1, xmax=10,
        ymin=0, ymax=100,
        xtick={1,2,3,4,5,6,7,8,9,10},
        ytick={0,20,40,60,80,100},
        legend cell align={left},
        % legend pos=outer north east,
        legend style={at={(1,0.73)},anchor=north east,nodes={scale=0.5, transform shape}},
        legend columns=1, 
        xmajorgrids=true,
        ymajorgrids=true,
        grid style=dashed,
    ]
    % \addlegendimage{line width=0.3mm,color=black};
    % \addlegendimage{line width=0.3mm,color=black,dashed};
    \addplot[color=mycolor1,mark=square,]
        coordinates {(1,80.45740740737102)(2,90.62433862429798)(3,92.2672839505755)(4,94.4629629629285)(5,95.507122507079)(6,95.89814814810407)(7,96.2802175190625)(8,96.601429910879)(9,96.6828645100351)(10,96.80952380947933)};\label{hwplot1}
        \addlegendentry{Proposed of detection}
    \addplot[color=mycolor2,mark=triangle,]
        coordinates {(1,72.35594421290733)(2,79.46409175281487)(3,86.26136850187319)(4,88.79017346914315)(5,90.36792021711737)(6,91.40647753871668)(7,92.08890888196801)(8,92.9413635985429)(9,94.21875551149377)(10,94.46802104281755)};\label{hwplot2}
        \addlegendentry{Proposed of exclusion}
    \addplot[color=mycolor3,mark=+,]
        coordinates {(1,2.855276782158441)(2,5.60403806584353)(3,8.411685328739443)(4,10.898816267247435)(5,13.10855814662375)(6,15.22181688125864)(7,16.74756873437818)(8,17.941659723799747)(9,19.37261137949504)(10,20.816202865002237)};\label{hwplot3}
        \addlegendentry{Clustering-based detection}
    % \addplot[color=mycolor4,mark=o,]
    %     \addlegendentry{Clustering-based exclusion}
    \addplot[color=mycolor5,mark=o,]
        coordinates {(1,7.644444444449705)(2,12.24382716048893)(3,16.832870370364833)(4,20.579629629622798)(5,23.861111111102382)(6,26.61555555554796)(7,29.69259259258209)(8,32.377330779042995)(9,34.36516754848934)(10,36.37522045853938)};\label{hwplot4}
        \addlegendentry{ECOD-based detection}
    \addplot[color=mycolor1,mark=square,dashed]
        coordinates {(1,73.7672839505856)(2,84.25767195763414)(3,86.74166666662708)(4,88.30497685181271)(5,91.94220945078846)(6,92.53098765427883)(7,93.02074829927726)(8,93.40823361819101)(9,93.7711396010969)(10,94.13859953699425)};
    \addplot[color=mycolor2,mark=triangle,dashed]
        coordinates {(1,66.80107574417546)(2,72.63566108407348)(3,78.96234652799431)(4,81.90400870745432)(5,83.49921291907378)(6,84.53057215742493)(7,85.53897791042578)(8,86.37893746846126)(9,88.29492986088382)(10,89.52834783429149)};
    \addplot[color=mycolor3,mark=+,dashed]
        coordinates {(1,6.277057613168598)(2,9.399770723103822)(3,11.2960204885734)(4,12.715619671236292)(5,14.193083983105603)(6,15.699283248493)(7,17.062093072899578)(8,18.158000232936916)(9,19.422073011200625)(10,20.781758313559165)};
    \addplot[color=mycolor5,mark=o,dashed]
        coordinates {(1,7.711111111110011)(2,12.790972222216937)(3,17.877777777771584)(4,21.095308641967273)(5,24.635185185175634)(6,27.698765432110023)(7,31.562962962959706)(8,33.94185185183898)(9,35.53535353534087)(10,38.2976190476043)};
    \legend{};
    % \addlegendentry{Solid lines for $\sigma_\mathrm{adv}=2$}
    % \addlegendentry{Dashed lines for $\sigma_\mathrm{adv}=4$}
    \end{axis}
    \end{tikzpicture}
    \caption{Attack (i)}
    \label{fig:resdetexc1}
    \end{subfigure}
    \hfill
    \begin{subfigure}[b]{0.66\columnwidth}
    \centering
    \begin{tikzpicture}%[scale=.75,font=\large]
    \begin{axis}[
        width=\columnwidth,
        xlabel={False positive rate ($P_{\text{FP}}$) [\%]},
        ylabel={True positive rate ($P_\text{TP}$) [\%]},
        xmin=1, xmax=10,
        ymin=0, ymax=100,
        xtick={1,2,3,4,5,6,7,8,9,10},
        ytick={0,20,40,60,80,100},
        legend cell align={left},
        % legend pos=outer north east,
        legend style={at={(1,0.73)},anchor=north east,nodes={scale=0.5, transform shape}},
        legend columns=1, 
        xmajorgrids=true,
        ymajorgrids=true,
        grid style=dashed,
    ]
    \addplot[color=mycolor1,mark=square,]
        coordinates {(1,71.39691358021556)(2,79.90952380948854)(3,81.8347826086585)(4,84.19259259255732)(5,85.61587301583411)(6,86.33635265696576)(7,87.03315412182455)(8,87.72491039422582)(9,88.21422558918569)(10,88.68442622946813)};
        \addlegendentry{Proposed of detection}
    \addplot[color=mycolor2,mark=triangle,]
        coordinates {(1,66.05915368365571)(2,69.88341967842185)(3,74.24887972588288)(4,76.4487788054074)(5,77.51903137502035)(6,78.61041769162631)(7,79.34771711586209)(8,79.79891665349589)(9,80.08884355075423)(10,81.02739094128432)};
        \addlegendentry{Proposed of exclusion}
    \addplot[color=mycolor3,mark=+,]
        coordinates {(1,7.717283950617107)(2,9.856193078323969)(3,11.405678032529598)(4,12.765826909819808)(5,14.168813741277214)(6,15.750957854405773)(7,17.09469993659278)(8,18.234298519001285)(9,19.479870449004844)(10,20.836066645741338)};
        \addlegendentry{Clustering-based detection}
    % \addplot[color=mycolor4,mark=o,]
    %     \addlegendentry{Clustering-based exclusion}
    \addplot[color=mycolor5,mark=o,]
        coordinates {(1,8.340740740744752)(2,13.207407407402527)(3,18.187037037030876)(4,21.950771604930736)(5,24.733333333327903)(6,28.231018518510315)(7,30.3930555555504)(8,32.88946360151977)(9,35.415740740730595)(10,38.050135501339544)};
        \addlegendentry{ECOD-based detection}
    \legend{};
    \end{axis}
    \end{tikzpicture}
    \caption{Attack (ii)}
    \label{fig:resdetexc2}
    \end{subfigure}
    \hfill
    \begin{subfigure}[b]{0.66\columnwidth}
    \centering
    \begin{tikzpicture}%[scale=.75,font=\large]
    \begin{axis}[
        width=\columnwidth,
        xlabel={False positive rate ($P_{\text{FP}}$) [\%]},
        ylabel={True positive rate ($P_\text{TP}$) [\%]},
        xmin=1, xmax=10,
        ymin=0, ymax=100,
        xtick={1,2,3,4,5,6,7,8,9,10},
        ytick={0,20,40,60,80,100},
        legend cell align={left},
        % legend pos=outer north east,
        legend style={at={(1,0.73)},anchor=north east,nodes={scale=0.5, transform shape}},
        legend columns=1, 
        xmajorgrids=true,
        ymajorgrids=true,
        grid style=dashed,
    ]
    \addplot[color=mycolor1,mark=square,]
        coordinates {(1,65.41527777774998)(2,75.21527777774498)(3,77.69938271601434)(4,79.42592592589003)(5,82.54861111108418)(6,83.54884259255504)(7,84.97173353905569)(8,85.22489711930255)(9,85.4780606995494)(10,85.79108796292418)};
        \addlegendentry{Proposed of detection}
    \addplot[color=mycolor2,mark=triangle,]
        coordinates {(1,58.31821512338305)(2,63.518635655666)(3,66.77162360407901)(4,69.51563119070525)(5,72.58027398084747)(6,74.16191266927143)(7,74.88805604644243)(8,75.6302659272425)(9,76.4123372523433)(10,76.95560592935599)};
        \addlegendentry{Proposed of exclusion}
    \addplot[color=mycolor3,mark=+,]
        coordinates {(1,4.422722722722639)(2,8.374279835390763)(3,10.66692341308837)(4,12.38097799848235)(5,14.010130273422416)(6,15.69462423588601)(7,17.079300286058027)(8,18.276372924648362)(9,19.59432946017871)(10,20.93697271922305)};
        \addlegendentry{Clustering-based detection}
    % \addplot[color=mycolor4,mark=o,]
    %     \addlegendentry{Clustering-based exclusion}
    \addplot[color=mycolor5,mark=o,]
        coordinates {(1,7.477777777779641)(2,12.43989898989382)(3,16.622592592590732)(4,21.025146198822237)(5,24.648148148139054)(6,27.965185185178076)(7,31.01172839505656)(8,34.029450418146495)(9,35.98074074072908)(10,38.530792420311943)};
        \addlegendentry{ECOD-based detection}
    \legend{};
    \end{axis}
    \end{tikzpicture}
    \caption{Attack (iii)}
    \label{fig:resdetexc3}
    \end{subfigure}
    \caption{Detection and exclusion $P_\text{TP}$ of the proposed and baseline methods. \ref{hwplot1} is detection and \ref{hwplot2} is exclusion performance for our scheme; \ref{hwplot3} for the clustering-based detection and \ref{hwplot4} for ECOD-based detection.}
    \label{fig:resdetexc}
\end{figure*}

\subsection{Detection Accuracy}
Detection accuracy measures the ability of a system to correctly identify the presence of rogue \acpl{ap} while maintaining a predefined maximum $P_{\text{FP}}$. It quantifies the proportion of detection outcomes that correctly raised an alarm that rogue \acpl{ap} were present. In other words, detection accuracy evaluates how effectively the system identifies potential threats without generating false positives. The comparison results are shown in Fig.~\ref{fig:resdetexc}. Solid lines are for $\sigma_\mathrm{adv}=2$ and dashed lines are for $\sigma_\mathrm{adv}=4$ in attack (i). 

% For attack (i) data, the proposed methods have much better true positive rates, from 67\% to 97\%. clustering-based detection exhibits relatively low true positive rates (less than 21\%) compared to the other methods across all false alarm rates and $\sigma_\mathrm{adv}$, indicating its limited effectiveness. ECOD-based detection shows improvements over clustering-based detection but still lags behind our proposed approaches, especially at lower false alarm rates. While ECOD-based detection demonstrates moderate performance, it may struggle to achieve high detection accuracy under challenging conditions. 

% For attack (ii) data, the proposed detection method demonstrates a consistent increase in true positive rates, ranging from approximately 71\% to 89\%, as false positive rates escalate from 1\% to 10\%. In contrast, clustering-based detection yields notably lower true positive rates, ranging from approximately 8\% to 21\%, while ECOD-based detection performs marginally better with rates ranging from approximately 8\% to 38\%. 

% For attack (iii) data, the proposed detection method achieves true positive rates ranging from approximately 65\% to 85\%, depending on the false alarm rate. In contrast, clustering-based detection only achieves true positive rates ranging from about 4\% to 21\%, and ECOD-based detection achieves rates ranging from approximately 7\% to 39\%. These results also clearly demonstrate the superior performance of the proposed approach.

In attack (i) scenarios (recall: attacks (i)--(iii) defined in Sec.~\ref{subsec:attsim}), our detection consistently achieves higher $P_\text{TP}$ (67\% to 97\%) compared to the clustering-based and ECOD-based detection methods. Clustering-based detection exhibits low rates (below 21\%) across all $P_{\text{FP}}$ and $\sigma_\mathrm{adv}$. While ECOD-based detection shows some improvement over clustering-based detection, it still falls short of our proposed approach variants, particularly at lower $P_{\text{FP}}$. In attack (ii) scenarios, our detection demonstrates increasing $P_\text{TP}$ (71\% to 89\%) as $P_{\text{FP}}$ grows, outperforming clustering-based and ECOD-based detection consistently. Similarly, in attack (iii), our detection achieves higher $P_\text{TP}$ (65\% to 85\%) compared to clustering-based and ECOD-based detection, across $P_{\text{FP}}$. 

The signals of randomly chosen \ac{ap} in attack (i) are manipulated with varying levels of additive noise characterized by $\sigma_\mathrm{adv}$: the detection accuracy tends to decrease as $\sigma_\mathrm{adv}$ increases. This trend suggests that higher levels of noise significantly affect the ability to accurately identify rogue \acpl{ap}, resulting in lower $P_\text{TP}$ and increased false positives. In attack (ii), where \acpl{rssi} of some \acpl{ap} are replaced by random values within a specified range, the detection accuracy is lower compared to attack (ii). This suggests that attack (ii) poses more of a challenge to detection algorithms compared to the noise-induced variations in attack (i). In attack (iii), $P_\text{TP}$ for our detection method is comparable to those for attack data (i) and (ii) at higher $P_{\text{FP}}$. A lower $P_{\text{FP}}$, the performance of detecting attack (iii) tends to be slightly lower compared to (i) and (ii). Despite this, the proposed detection still outperforms clustering-based detection and ECOD-based detection across all attacks. %It is essential to note that the effectiveness of detection algorithms in mitigating different types of attacks depends on various factors, including the specific characteristics of the Wi-Fi network and the sophistication of the attack strategies. 

\subsection{Rogue \ac{ap} Exclusion}
Rogue \ac{ap} exclusion requires correctly identifying those in the subsets deemed affected. Exclusion accuracy measures the ability to exclude rogue \acpl{ap} after detection of an attack (rogue \ac{ap} in range). The results are also shown in Fig.~\ref{fig:resdetexc}. For attack (i), the proposed exclusion achieves $P_\text{TP}$ ranging from 67\% to 94\%. While its $P_\text{TP}$ is slightly lower than the proposed detection naturally, as it is challenging to correctly exclude rogue \acpl{ap}, it still has a high accuracy. Similarly, in attack (ii) scenarios, the exclusion method achieves $P_\text{TP}$ ranging from approximately 66\% to 81\% for $P_{\text{FP}}$ of 1\% to 10\%, showing improvement with increasing $P_{\text{FP}}$. However, for attack (iii) scenarios, the exclusion accuracy initially drops to around 60\% at lower $P_{\text{FP}}$ due to the difficulty in distinguishing rogue and legitimate \acpl{ap} in close proximity. As $P_{\text{FP}}$ increases, $P_\text{TP}$ improves to 77\%.

% For attack (i) data, the proposed exclusion method achieves true positive rates between 67\% and 94\%, with corresponding false positive rates ranging from 1\% to 10\%. Since not all detected $t$ can exclude the right rogue \ac{ap}, its true positive rate is lower than the proposed detection. However, it is still better than other baseline detections. 

% For attack (ii) data, the true positive rates of the proposed exclusion method range from approximately 66\% to 81\% for false positive rates ranging from 1\% to 10\%. Similar to the proposed detection method, the exclusion method also demonstrates an improvement in true positive rates with increasing false positive rates. 

% For attack (iii) data, the exclusion accuracy drops to over 60\% at lower false alarm rates. This decrease can be attributed to the difficulty of differentiating between rogue and legitimate \acpl{ap} that are in close proximity to each other. As the false alarm rate increases, the true positive rate improves to 77\%. Although the rates are slightly lower compared to the detection method, the exclusion still has a larger performance gain than other methods. 

\begin{figure}
    \centering
    \includegraphics[width=\columnwidth]{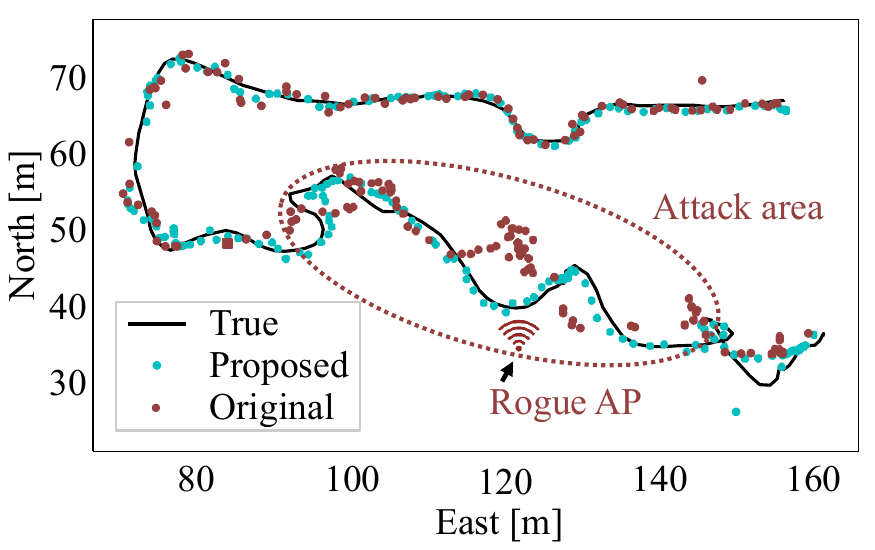}
    \caption{An illustration of the positioning results before (i.e., original) and after (i.e., proposed) exclusion.}
    \label{fig:recpos}
\end{figure}
\subsection{Position Recovery}
Position recovery refers to the process of determining the position of a Wi-Fi client based on \acpl{rssi} from the deemed benign \acpl{ap} after the estimated rogue \acpl{ap} were excluded. Then, the indoor Wi-Fi positioning algorithm relying solely on benign \ac{rssi} measurements estimates the client position accurately. The effectiveness of position recovery can be evaluated by comparing the reference positions (ground truth) with the positioning results obtained before and after the exclusion of rogue \acpl{ap}. This comparison allows to assess the accuracy of the client's position estimation and, naturally, how rogue \ac{ap} detection and exclusion improve the accuracy of the Wi-Fi positioning system. The positioning results before and after rogue \acpl{ap} exclusion are shown in Fig.~\ref{fig:recpos}. The red solid dots are the original Wi-Fi positioning results, showing the attack area deviation from the actual path, drawn as a black line. In contrast, the cyan-blue solid dots are the results of the proposed method with much lower deviation. 

\subsection{Effect of Sampling}
Based on the sampling strategy for subsets proposed in Sec.~\ref{subsec:samstr}, we evaluate the detection $P_\text{TP}$ for different ratios of sampling subsets, under attack (i) with $\sigma_\mathrm{adv}=2$. $P_{\text{FP}}$ is fixed to 0.05. $P_\text{TP}$ versus sampling ratio is shown in Tab.~\ref{tab:effsam}. The results demonstrate a clear trend of increasing $P_\text{TP}$ with higher sampling ratios. As the sampling ratio increases from 0.25 to 1.0, there is an improvement in $P_\text{TP}$ from 83\% to 96\%. This shows that a higher sampling ratio enables capturing a wider range of \ac{rssi} subsets; while with a lower sampling ratio, the algorithm may miss crucial subsets, leading to decreased accuracy. However, accuracy is not very sensitive to the sampling ratio. When the sampling rate decreases from 1 to 0.25 (4 times computation reduction), $P_\text{TP}$ only increases by 13\%. While higher sampling ratios generally lead to better performance, there may be practical considerations, computational resources and time constraints to take into account.  
\begin{table}
% \normalsize
\centering
\caption{$P_\text{TP}$ versus sampling ratio.}
\begin{tabular}{lrrrr}
\hline
\hline
Sampling Ratio & 0.25 & 0.4 & 0.7 & 1.0 \\ %0.1 & 
\hline
Detection Accuracy & 83\% & 89\% & 95\% & 96\% \\ %& 46\% 
\hline
\hline
\end{tabular}
\label{tab:effsam}
\end{table}

\section{Limitations}
A limitation of this work emerges when the attacker (rogue \acpl{ap}) can adjust the transmission power so that it can appear to be close to the victim \acpl{ap}. If the position computed including rogue \ac{ap} measurements is very close to that based on benign \acpl{ap}, they are clustered together and consequently the rogue \ac{ap} is not excluded. In addition, we have not considered our approach in conjunction with other schemes that reveal inconsistencies or spurious traffic. 

\section{Conclusions}
This paper proposes a scheme that effectively identifies rogue \acpl{ap} without the need for specialized hardware. We rely on redundant position information obtained from Wi-Fi positioning, through a \ac{raim}-style method for cross-validation. The two key components of our approach, subset generation and position validation using Gaussian mixture \ac{raim}, show higher rogue \ac{ap} detection accuracy compared to baseline methods. In future work, we will consider integrating our position-based detection with other cross-validation approaches. We will combine the sampling strategy with pruning algorithms to reduce computations. We will compare to other related solutions (as baselines), consider more realistic attacks, and work on locating rogue \acpl{ap}. 

\section*{Acknowledgment}
This work was supported in parts by the Knut and Alice Wallenberg Foundation and the China Scholarship Council. 

\bibliographystyle{IEEEtran}
\bibliography{reference/references}

\end{document}